\documentclass[prl,aps,showpacs,floatfix,floats,nobalancelastpage,twocolumn]{revtex4}
\usepackage{graphicx,bm}    

\begin{document}
\title{Semiclassical methods for multi-dimensional systems bounded by finite potentials}
\author{Wen-Min Huang$^1$, Cheng-Hung Chang$^{2,3}$, Chung-Yu
Mou$^{1,2}$}
\date{\today}
\affiliation{$^1$ Department of Physics, National Tsing-Hua
university, Hsinchu 300, Taiwan \\
$^2$ Physics Division, National Center for Theoretical Sciences,
Hsinchu 300, Taiwan \\
$^3$ Institute of Physics, National Chiao Tung university, Hsinchu
300, Taiwan}



\begin{abstract}
This work studies the semiclassical methods in multi-dimensional
quantum systems bounded by finite potentials. By replacing the
Maslov index by the scattering phase, the modified transfer
operator method gives rather accurate corrections to the quantum
energies of the circular and square potential pots of finite
heights. The result justifies the proposed scattering phase
correction which paves the way for correcting other semiclassical
methods based on Green functions, like Gutzwiller trace formula,
dynamical zeta functions, and Landauer-B\"uttiker formula.
\end{abstract}



\maketitle


Semiclassical approaches are techniques to study quantum systems
in classical limit \cite{Brack}. In these approaches, physical
quantities are expressed by the Green function of the system and
replaced by its underlying classical trajectories under
stationary phase approximation. Well known examples range over
Gutzwiller trace formula, dynamical zeta function, and transfer
operators, which are aimed at determining the quantized energies
of closed systems \cite{Brack}. For open systems, the most
prominent example might be the Landauer-B\"utticker formula for
charge current transport through open quantum dots
\cite{Jalabert}. Recently semiclassical approaches also have been
used to study spin current transport \cite{AM} and spin dynamics,
which clarifies the suppression of D'yakonov-Perel' spin
relaxation in mesoscopic quantum dots \cite{CHC}. All these
approaches can relate quantum problems to the ergodicity property
of their corresponding classical dynamics and gives more
transparent pictures to complex phenomena, including the
signatures of quantum chaos \cite{Brack}.

Besides this conceptual contribution, some semiclassical methods
provide efficient numerical techniques for less time-consuming
calculations. Their results are especially accurate in mesoscopic
systems in which the de Broglie wavelength of the Fermi electrons
is much shorter the sample size. The devices in this scale are
often fabricated by lithography or controlled by confining
potentials. A convenient theoretical approach to study these
quantum systems is assuming them to be bounded by infinite
potential. This largely simplifies the formulism of the
semiclassical methods, since the quantum particle has the same
phase change for arbitrary particle energy after it is reflected
by this potential. This phase change is carried in the well known
Maslov index. However, for real systems beyond this assumption,
the accuracy of the conventional semiclassical methods using
Maslov index become out of control. The usual way is going back
to solve the Schr\"odinger equation with a boundary combined with
Dirichlet and von Neuman conditions. But one could ask whether
the conventional semiclassical methods still work after taking
certain effective correction.

A natural candidate for this correction is an effective phase
change in the wave function after it is bounced back by the
potential. This phase can be understood as the dwell time of the
quantum particle penetrating into and staying inside the
potential barrier. Indeed, in the one-dimensional (1D) potential
well, it has been shown that replacing the Maslov index in the
Green function by a scattering phase, this function gives an
exact quantization rule identical with the
Wentzel-Kramers-Brillouin (WKB) method \cite{HHLin}. This
relation stirs up the motivation how to extend this result to
multi-dimensional systems. The current paper demonstrates this
extension in the example of Bogomolny's transfer operator (BTO)
method \cite{EB} and tested it in a circular billiard and a
square billiard bounded by potential pots of finite hight. The
calculated energies are surprisingly accurate, which justifies
the suggested scattering phase correction for multi-dimensional
systems. Although the result is presented in the BTO method, it
should be valid for all semiclassical methods as long as they are
based on the Green function.


The WKB method might be the simplest semiclassical method. For a
particle of mass $m$ and energy $E$ bounded by an 1D potential
$V(x)$, the solution of its Schr\"{o}dinger equation can be
approximated by the WBK wave function \cite{Landau},
\begin{equation}
\psi(x)=\frac{1}{\sqrt{p(x)}}{\rm exp}\left[ \pm
\int_{x_i}^{x}p(x')dx' \right],
\end{equation}
with $p(x)=\sqrt{2m[E-V(x)]}$, if the de Broglie wavelength
$\lambda(x)=2\pi\hbar/p(x)$ varies slowly compared to the
potential. This requirement is usually violated at the classical
turning point $x_0$, where the momentum changes sign. At that
point one has $E=V(x_0)$ which gives rise to a vanishing $p(x_0)$
and a singular function $\psi(x)$. If the potential varies slowly
around $x_0$, the exponentially decreasing real wave function
outside this point should be associated with the oscillating wave
function inside this point. This enforces the incident wave
function to take a $scattering$ $phase$ after reflection. This
phase is equal to $\pi/2$ in the semiclassical (short wave) limit
\cite{Landau}. If the particle is reflected back and forth
between two turning points $x_1$ and $x_2$ on two potential
barriers, the particle should take two scattering phases
$\phi_1=\phi_2=\pi/2$ during one period of motion. The total
phase then equals $2n\pi$ with an integer $n$, which leads to the
WKB quantization condition \cite{Landau}
\begin{equation}\label{wkb1}
\frac{1}{\hbar}\oint p(x)dx=\frac{2}{\hbar}\int^{x_2}_{x_1}
p(x)dx=2\pi\left(n+\frac{\mu}{4}\right),
\end{equation}
where the Maslov index $\mu=2$ corresponds to the two reflections
during one period of particle motion.

If the potential does not vary sufficiently slowly at the turning
points, for example the step function barrier, the phase change
$\pi/2$ is no longer a good approximation. A general scattering
phase should be determined quantum-mechanically \cite{HF},
\begin{equation}\label{wkb2}
\frac{1}{\hbar}\oint p(x)dx=2n\pi+\phi_1(E) + \phi_2(E),
\end{equation}
where the phase changes $\phi_1(E)$ and $\phi_2(E)$ at the
turning points $x_1$ respectively $x_2$ become a function of $E$.
As an example, let this particle move in an 1D finite square well
with $V(x)=0$ for $0<x<L$ and $V_0$ otherwise, where $L$ is the
well width and $V_0>0$ is the potential height. Solving the
Schr\"odinger equation with the continuous boundary condition,
the scattering phase can be calculated,
\begin{equation}\label{sp1}
\phi_s(E)=\cos^{-1}\left[2\left(\frac{E}{V_0}\right)-1\right].
\end{equation}
Substituting Eq. (\ref{sp1}) into $\phi_1(E)$ and $\phi_2(E)$ in
Eq. (\ref{wkb2}), one obtains an exact quantization rule for the
1D finite square well.

For general $k$-dimensional systems with $k\geq 2$, many
quantization rules have been derived to extend the WKB method,
including Gutzwiller trace formula, dynamical zeta function, and
transfer operators \cite{Brack}. All of them are based on the
semiclassical Green function
\begin{eqnarray}
G({\bf r},{\bf r'};E)=\frac{1}{(2\pi
i\hbar)^{k/2}}\sum_\gamma\sqrt{\left|\det\frac{\partial S_{\rm
cl}({\bf r},{\bf r'};E)}{\partial {\bf r}\partial {\bf
r}'}\right|} \nonumber \\
\hspace{1cm}\times\exp\left[\frac{i}{\hbar}S_{\rm cl}({\bf r},{\bf
r'};E)-i\mu\frac{\pi}{2}\right], \label{gf1}
\end{eqnarray}
which is a sum over all classical trajectories $\gamma$ starting
from ${\bf r'}$ and ending at ${\bf r}$ \cite{VV,Brack}. The
function $S_{\rm cl}({\bf r},{\bf r'};E)$ therein is the action
of the particle from $\bf r'$ to $\bf r$ along $\gamma$. The
Maslov index $\mu$ stands for the total number of the turning
points between ${\bf r'}$ and ${\bf r}$ along $\gamma$. For 1D
finite square well, the quantization rule derived from Eq.
(\ref{gf1}) with the replacement $\mu\frac{\pi}{2}$ by $\phi_s$
in Eq. (\ref{sp1}) is identical with the quantization rule of Eq.
(\ref{wkb2}) with the same phase correction $\phi_s$
\cite{HHLin}.

\begin{figure}[htbp!]
\begin{center}
\includegraphics[width=7.6cm]{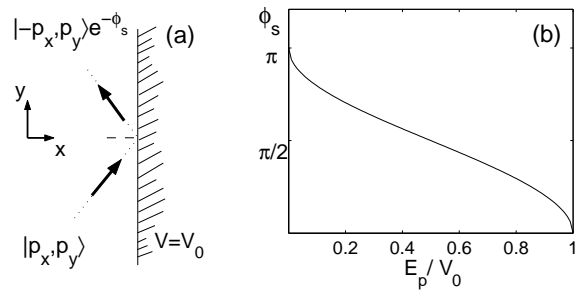}
\caption{(a) The incident state $|p_x,p_y \rangle$ and the
reflected state $|-p_x,p_y\rangle\,e^{-\phi_s}$ on the confining
potential $V_0$. (b) The scattering phase $\phi_s$ of the
reflected state is a function of $E_p/V_0$.}\label{SP}
\end{center}
\end{figure}

For multi-dimensional systems, intuitively, the magnitude of the
scattering phase should depend on the incident angle of the
particle wave. Naively one could suggest the phase $\phi_s$ in Eq.
(\ref{sp1}) to be only related to the perpendicular component of
the incident wave with respect to the boundary. Figure \ref{SP}(a)
shows an example of a 2D step potential $V_c(x,y)=V_0 \Theta(x)$
with the potential high $V_0>0$ for $x\geq 0$ and $0$ for $x<0$,
where $\Theta(x)$ is the Heviside function. The momentum $p_x$
perpendicular to the boundary has energy $E_p=p_x^2/2m$, which
will replace the total energy $E$ in Eq. (\ref{sp1}) for
multi-dimensional systems. This phase is a value between $0$ and
$\pi$, depending on the ratio $E_p/V_0$, as shown in Fig.
\ref{SP}(b). If the potential barrier is high, that is
$E_p/V_0\ll1$, the scattering phase will approach $\pi$, which is
the same as the infinitely high potential barrier.

Next, the scattering phase for the perpendicular wave component
discussed above will be merged into BTO method. Let us consider
again the particle of mass $m$ and energy $E$ moving in a
$k$-dimensional system. Select a Poincar\'{e} section (PS)
$\Sigma$ in the configuration space of this system, such that
almost all classical trajectories pass this section \cite{good1}.
The original transfer operator ${\cal T}(E)$ is defined as the
integral operator \cite{EB}
\begin{equation}\label{transfer1}
{\cal T}(E)\psi(q)=\int_{\Sigma}T(q,q';E)\psi(q')dq',
\end{equation}
acting on some function $\psi(q')$ on $\Sigma$. The integral
kernel,
\begin{eqnarray}\label{transfer2}
T(q,q';E)\hspace{-0.1cm}&=&\hspace{-0.2cm}\sum_\gamma\frac{1}{(2\pi
i\hbar)^{(k-1)/2}}\sqrt{\left|{\rm det}\frac{\partial
S_{\rm cl}(q,q';E)}{\partial q\partial q'}\right|}\nonumber \\
& &\times\hspace{0.05cm}{\rm exp}\left[\frac{i}{\hbar}S_{\rm
cl}(q,q';E) -i\nu\frac{\pi}{2}\right],
\end{eqnarray}
is defined as the sum over all possible classical trajectories
$\gamma$'s from the initial point $q'\in \Sigma$ to the final
point $q \in \Sigma$. The action $S_{\rm cl}(q,q';E)$ is the same
as before in Eq. (\ref{gf1}). The Maslov index $\nu$ counts the
number of the crossing points of $\gamma$ through $\Sigma$ from
the same side of $\Sigma$. According to the BTO method, the zeros
of the Fredholm determinant $|\det(1-{\cal T}(E))|$ of the
transfer operator ${\cal T}(E)$ are the energy eigenvalues of the
quantum system.

The kernel in Eq. (\ref{transfer2}) is derived from the
semiclassical Green function in Eq. (\ref{gf1}), which can be
divided into two parts, $G(q,q';E)=G^{\rm
osc}(q,q';E)-G_0(q,q';E)$, where $G^{\rm osc}(q,q';E)$ is the
contribution from long trajectories and $G_0(q,q';E)$ is the
contribution form short trajectories. After coordinate deduction
from $k$-dimensional space to ($k-1$)-dimensional space on
$\Sigma$ \cite{EB}, the quantization condition
$|\det(G(q,q';E))|=0$ can be expressed as $|\det(1-{\cal
T}(E))|=0$, where the identity operator comes from $G_0$ and the
operator ${\cal T}(E)$ originates from $G^{\rm osc}$. The entire
derivation holds the same when the phase in $G(q,q';E)$ is
modified. The transfer operator ${\cal T}_m(E)$ modified with the
scattering phase for perpendicular wave component then has the
kernel
\begin{eqnarray}\label{transfer3}
T_m(q,q';E)\hspace{-0.1cm}&=&\hspace{-0.2cm}\sum_\gamma\frac{1}{(2\pi
i\hbar)^{(k-1)/2}}\sqrt{\left|{\rm det}\frac{\partial
S_{\rm cl}(q,q';E)}{\partial q\partial q'}\right|}\nonumber \\
& &\times\hspace{0.05cm}{\rm exp}\left[\frac{i}{\hbar}S_{\rm
cl}(q,q';E) -i\phi_s(E_p)\right].
\end{eqnarray}
The efficiency of this modified operator will be tested in the
following two 2D integrable systems.


\begin{figure}[htbp!]
\begin{center}
\includegraphics[width=8.5cm]{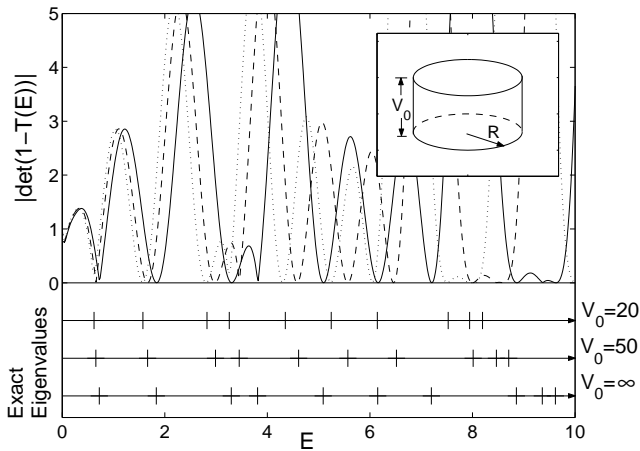}
\caption{The upper part shows the Fredholm determinant
$|\det(1-{\cal T}_m(E))|$ of the modified transfer operator
${\cal T}_m(E)$ for the potential pot in the inset for $V_0=20$,
$50$, and $\infty$ with $R=2$. The points on the three bottom
lines depict the exact quantum energies for $V_0=20$, $50$, and
$\infty$ with $R=2$.}\label{circleplot}
\end{center}
\end{figure}

The first system is a 2D circular quantum dot bounded by a finite
potential, as shown in the inset of Fig. \ref{circleplot}. The
potential pot has radius $R$ and height $V_0$, that is $V(r)=V_0$
for $r\geq R$ and $V(r)=0$ for $r< R$. Analytically the energy
eigenvalues of its Schr\"odinger equation can be calculated by
matching the boundary conditions at radius $r=R$ \cite{math}.
Setting Planck constant $\hbar=1$, the mass $m=1$, and the pot
radius $R=2$, the exact energy eigenvalues for different
potential height $V_0=20$, $50$, and $\infty$ are denoted on the
three bottom lines of Fig. \ref{circleplot}. The dotted, dashed,
and solid curves in the upper part of Fig. \ref{circleplot} are
the Fredholm determinant $|\det(1-{\cal T}_m(E))|$ of the
modified transfer operator for $V_0=20$, $50$, and $\infty$. The
zeros of these functions determine the semiclassical quantum
energies of the systems.

\begin{figure}[htbp!]
\begin{center}
\includegraphics[width=8.5cm]{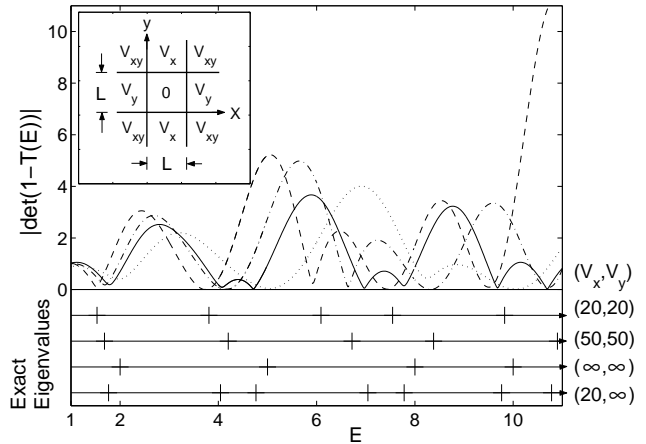}
\caption{The upper part shows the Fredholm determinant
$|\det(1-{\cal T}_m(E))|$ of the modified transfer operator ${\cal
T}_m(E)$ for the square potential pot in the inset with the
potential configuration $(V_x,V_y)=(20,20)$, $(50,50)$,
$(20,\infty)$, and $(\infty,\infty)$ where $V_{xy}=V_x+V_y$. The
points on the four bottom lines depict the exact quantum energies
for $(V_x,V_y)=(20,20)$, $(50,50)$, $(20,\infty)$, and
$(\infty,\infty)$.} \label{squareplot}
\end{center}
\end{figure}

The second system is the 2D square quantum dot with the confining
potential as shown in the inset of Fig. \ref{squareplot}. Therein
the $xy$-plane is separated into nine regions with different
constant potential heights $V_x$, $V_y$, and $V_{xy}$ at each
region, where $V_x>0$, $V_y>0$, and $V_{xy}=V_x+V_y$. This 2D
problem can be reduced to two independent 1D finite wells with
potential heights $V_x$ and $V_y$ and solved separately \cite{QM}.
Combining the eigenvalues of these two separated systems, the
total quantum energies of this 2D system for $(V_x,V_y)=(20,20)$,
$(50,50)$, $(20,\infty)$, and $(\infty,\infty)$ are determined and
depicted on the four bottom lines in Fig. \ref{squareplot}, where
$\hbar=m=1$ as before and the well length $L$ is normalized by the
condition $2\pi^2/L^2=1$. The upper part of this Figure shows the
Fredholm determinants $|\det(1-{\cal T}_m(E))|$ of the modified
BTO. The dashed, dash-dotted, dotted, and solid curves represent
these functions for the potential configurations
$(V_x,V_y)=(20,20)$, $(50,50)$, $(\infty,\infty)$, and
$(20,\infty)$.


Figure 2 and 3 shows surprisingly accurate quantum energies after
the scattering phase correction. Taking Fig. 2 as example, the
exact $10$-th energy has a remarkable left shift after the
potential height is reduced from $V_0=\infty$ to $V_0=20$. This
shows how large the error could be when simplifying a deep
potential pot, even with the large ratio $V_0/R=10$, by an
infinitely high pot. Without scattering phase correction, the
semiclassically calculated energies are the zeros of the solid
curve. The 10-th zero of this curve is close to the $10$-th point
on the bottom line of $V_0=\infty$, but far apart from the
$10$-th point for $V_0=20$. However, after taking the scattering
phase, the determinant (dotted curve) shifts leftward quite a lot
and its zeros largely approach the exact energies for $V_0=20$.

Quantitatively we can define values to characterize these errors.
Suppose $E_i$ is the $i$-th exact energy of a quantum system
bounded by some potential pot of height $V_0<\infty$ and
$\tilde{E}_i$ is the corresponding semiclassical energy
approximated by the modified BTO. Let the value
$\delta_i=\tilde{E}_i-E_i$ be the difference between these two
energies and the ratio $\Delta_i=\delta_i/E_i$ is the error of the
i-th eigenvalue. For the special case of $V_0=\infty$ the values
defined above are furnished with a superscript $\infty$ as
$E_i^\infty$, $\tilde{E}_i^\infty$, and
$\delta^\infty_i=\tilde{E}^\infty_i-E^\infty_i$ respectively. The
relative error $\Gamma_i$ of the $i$-th energy is then defined as
the ratio
\begin{equation}\label{RE}
\Gamma_i = \frac{\left|\delta_i -
\delta^\infty_i\right|}{|E_i-E_i^\infty|}.
\end{equation}
The denominator denotes the exact energy shift after the
potential height is reduced from $\infty$ to $V_0$. The numerator
represents the error $\delta_i$ for $V_0<\infty$ subtracted by
the basic semiclassical error $\delta_i^\infty$ from infinite
potential.

\begin{figure}[htbp!]
\begin{center}
\includegraphics[width=8.5cm]{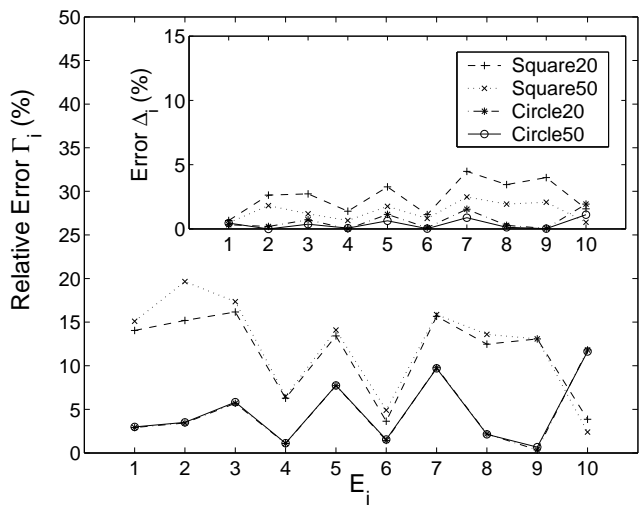}
\caption{The dotted and dashed curves in the main plot denote the
relative errors $\Gamma_i$ of the 2D circular potential pots with
potential height $50$ and $20$, as shown in the inset of Fig.
\ref{circleplot}. The dash-dotted and solid curves represent
$\Gamma_i$ for the 2D square potential pots of configuration
$(V_x,V_y)=(50,50)$ and $(20,20)$, as shown in the inset of Fig.
\ref{squareplot}. The corresponding errors $\delta_i$ of these
systems are shown in the inset.}\label{error}
\end{center}
\end{figure}

According to these definitions, the errors $\delta_i$ of the
lowest ten energies from Fig. \ref{circleplot} and
\ref{squareplot} are plotted in the inset of Fig. \ref{error} and
their corresponding relative errors $\Gamma_i$ are plotted in its
main figure. All $\delta_i$ are bounded by $5\%$ and all
$\Gamma_i$ are bounded by $20\%$. Roughly speaking, the
scattering phase in the modified BTO has corrected at least
$80\%$ of the energy error due to the potential reduction from
$\infty$ to $V_0$. This justifies the scattering phase correction
proposed above for finitely confined systems.

Notably, one cannot expect $100\%$ correction in this first order
correction. Remember that there still exists other degrees of
freedom in the system, which are not included in the scattering
phase. For instance, the modified transfer operator ${\cal
T}_m(E)$ is the same for square potential pots of different
$V_{xy}$, although they have different quantum energies. This
$V_{xy}$ difference can only be distinguished from higher order
corrections beyond the current scattering phase. Furthermore, if
the particle energy $E$ is close to the potential height $V_0$,
the particle can penetrate into the potential well quite long and
the wave property of the particle prevails its particle property.
In this regime, the correction deviation from the modified BTO
increases. However, that is not because this first correction is
wrong, but because higher order corrections are required in the
semiclassical approach.

Finally, almost all semiclassical methods are based on the Green
function of the quantum system. The successful result in the
modified BOT gives a clear direction for extending the quantum
correction to other semiclassical methods, including Gutzwiller
trace formula, dynamical zeta functions, and Landauer-B\"uttiker
formula.

We thank Hsiu-Hau Lin for fruitful discussions. This work was
supported by the National Science Council at Taiwan under Grant
Nos. NSC 93-2112-M-007-009.

\end{document}